\documentclass[preprint,samsmath,amssymb,floatfix,superscriptaddress]{revtex4}
\usepackage{graphicx}
\usepackage{dcolumn}
\begin{document}
\title{Locally preferred structure in simple atomic liquids}
\author{S.~Mossa and  G.~Tarjus} 
\affiliation
{Laboratoire de Physique Th\'eorique des Liquides,
Universit\'e Pierre et Marie Curie, 4 place Jussieu, 
Paris 75005, France}
\date{\today}
\begin{abstract}
We propose a method to determine the locally preferred structure 
of model liquids. This latter is obtained numerically as the 
global minimum of the effective energy surface of clusters formed
by small numbers of particles embedded in a liquid-like environment.
The effective energy is the sum of the intra-cluster interaction potential
and of an external field that describes the influence of the 
embedding bulk liquid at a mean-field level.
Doing so we minimize the surface effects present in isolated clusters
without introducing the full blown geometrical frustration present
in bulk condensed phases. We find that the locally preferred structure
of the Lennard-Jones liquid is an icosahedron, and that the liquid-like
environment only slightly reduces the relative stability of the icosahedral 
cluster.
The influence of the boundary conditions on the nature of the ground-state
configuration of Lennard-Jones clusters is also discussed.
\end{abstract}
\maketitle
\section{Introduction}
\label{intro}
A liquid is said to be {\em supercooled} when it is possible
to cool it below its melting temperature $T_m$ without 
crystallizing. The supercooled liquid phase is metastable 
with respect to the underlying crystal, and it is characterized 
by a dramatic increase of the viscosity and the relaxation times upon 
lowering the temperature, an increase that eventually leads to
glass formation.
These dramatic changes in the dynamic properties are not accompanied
by strong signatures in the structural quantities, such as the static
structure factor. Yet, it has been suggested that both supercooling
and glass formation were deeply connected to the structure of the liquid,
more precisely to a competition between extension of a local liquid
order, different than that of the crystal, and global constraints 
associated with tiling of the entire space~\cite{sadoc99,nelson89,kivelson95}.
This competition has been termed 
{\em geometric (or topological) frustration}~\cite{sadoc99,nelson89}.

Some fifty years ago, Frank put forward the following argument to explain
supercooling of liquids~\cite{frank52}. If one considers atomic liquids
in which atoms interact via spherically symmetric potentials like the
Lennard-Jones potential, the local arrangement of the atoms
that is preferred is not the clusters associated with crystalline order
(face-centered cubic and hexagonal close-packed lattices), but a 
polytetrahedral packing, the icosahedron formed by $13$ atoms.
The energy of such an icosahedral arrangement interacting via the
Lennard-Jones potential is indeed $8.4\%$ lower than the close-packed
crystalline clusters.
Local icosahedral order should then be prevalent in liquids but, because 
of the $5$-fold rotational symmetry of the icosahedron, it cannot tile
the entire space and form a crystal: this is a manifestation of geometrical
frustration.
Crystallization then requires a rearrangement of the local structure of 
the liquid, which leads to a strong first-order freezing transition and 
allows supercooling of the liquid.
Since then, there has been a large body of experimental and simulation
work confirming the prevalence of local icosahedral, or more generally
polytetrahedral, order in atomic liquids and metallic 
glasses~\cite{nelson89,filipponi99,reichert00,shenk02,steinhardt83,jonsson88,kondo90,yonezawa91,shumway95,jund97,dzugutov02,doye03}.
The tendency to form icosahedral order has been shown to increase
as the temperature is lowered~\cite{nelson89,jonsson88,kondo90,yonezawa91,shumway95,jund97,dzugutov02,doye03}.

Frank's argument has, however, one shortcoming~\cite{doye01}: the
$13$-atom cluster considered is isolated, so that most of the energetics
is related to the surface, a situation that of course does not
occur in bulk liquids. How can one remedy this problem?
An {\em a priori} easy way would be to study directly the local
arrangement of the atoms in a bulk liquid. In such a case, however,
it is found that the proportion of icosahedra among all the
$13$-atom groups is very small, typically a few percent~\cite{kondo90}.
This is indeed to be expected since geometrical frustration is present, which
opposes the growth of icosahedral order and distorts the local polytetrahedral
arrangements~\cite{nelson89,sachdev92}.
How to disentangle then the determination of the locally preferred 
structure from frustration effects?
The method we propose in this work is to consider the influence
of the bulk liquid on a given $13$-atom cluster at some mean-field level,
so that surface effects can be reduced and made more realistic for describing
a condensed phase, whereas geometrical frustration is strongly inhibited.
The main advantage of this method is that it can be extended to study the
locally preferred structure of molecular liquids, for which a priori
topological arguments do not easily provide the symmetries of all possible
local arrangements of the molecules, nor the nature of geometrical frustration.

More specifically, in this article we consider the ground-state of a cluster of
$13$ atoms interacting via a Lennard-Jones potential (smaller and larger 
clusters are also considered); the atoms are placed in a cavity and are subject
to an external field that mimics the interaction with the rest of the liquid.
The structure of the outside liquid only enters the calculation
via the bulk pair distribution function (known from previous simulation 
studies).
By means of an optimization algorithm~\cite{wales97,doye98}
we find that the ground-state of the cluster, i.e., the global minimum
of the (effective) energy surface formed by the intra-cluster interactions
and the external field, is of icosahedral symmetry, therefore generalizing
Frank's result. For sake of comparison we consider in addition other
boundary conditions for the cluster. These conditions describe different
types of environments: free boundary conditions (isolated cluster), 
periodic boundary conditions (periodic tiling of space), and 
icosahedral-like boundary conditions (hypothetical non-frustrated system).
\section{Method and choice of the boundary conditions}
\label{method}
We consider a system of atoms interacting via a pair-wise additive 
spherically symmetric potential $v(r)$, where $r$ is the 
center-to-center distance. A number $N$ of atoms are placed
in a spherical cavity $\cal C$ of radius $R_C$, that we
envisage as surrounded by bulk liquid made of the same atoms,
and characterized by the temperature $T$ and the density $\rho$.
As explained above, we do not want to fully account for the
liquid structure because geometrical frustration would obscure
the nature of the local order. We rather resort to a mean-field 
type of description in which the liquid outside the cavity
is considered as a continuum, characterized by a (known) 
pair distribution function $g(r;T,\rho)$
that is not affected by the fact that a cavity has been carved out.
The potential energy acting on a given atom at position ${\bf r}$
inside the cavity due to the outside liquid is thus described as
\begin{equation}
W({\bf r};R_C)=\frac{\rho}{2}\int_{{\bf r'}\not\in {\cal C}} 
d^3 {\bf r'}\; g(|{\bf r}-{\bf r'}|) v(|{\bf r}-{\bf r'}|),
\label{meanfield:eq}
\end{equation}
where the integral is over all positions outside the cavity.
Taking the center of the cavity as origin and transforming to
spherical coordinates, one finds after standard manipulations
\begin{equation}
W(r;R_C)=\pi\rho\left\{\int_{R_C-r}^{R_C+r}dx\; x^2 g(x) v(x)
\left[1-u(x;r,R_C)\right]+2 \int_{R_C+r}^{\infty}dx\; x^2 g(x) v(x)
\right\},
\label{pot_liq:eq}
\end{equation}
where
\begin{equation}
u(x;r,R_C)=\left(\frac{R_C^2-r^2-x^2}{2 r x}\right).
\label{u_def:eq}
\end{equation}
Note that $u(x=R_C-r)=-u(x=R_C+r)=1$. The detail of the calculation is 
given in the Appendix~\ref{calcul:app}.

In view of the implementation of the optimization 
algorithm~\cite{wales97,doye98}, we need an explicit expression for the 
first derivatives of the external potential, both with respect to $R_C$ 
and $r$; we obtain
\begin{equation}
\frac{\partial}{\partial R_C} W(r;R_C)\bigg|_{r}=
-\pi \rho \left(\frac{R_C}{r}\right) \int_{R_C-r}^{R_C+r}
dx\; x g(x) v(x),
\label{der_RC:eq}
\end{equation}
and
\begin{equation}
\frac{\partial}{\partial r}W(r;R_C)\bigg|_{R_C}=
\pi \rho \int_{R_C-r}^{R_C+r} dx\; x^2 g(x) v(x)
\left(\frac{R_C^2+r^2-x^2}{2 x r^2}\right).
\label{der_r:eq}
\end{equation}
The second derivatives, needed for the calculation of the Hessian
matrix and the study of the transition states~\cite{wales97,doye98},
can be obtained in a similar way.

The total potential energy for the $N$ atoms of the
embedded cluster is the sum of the atom-atom interaction
potentials inside the cavity and of the external potential,
\begin{equation}
{\cal U}\left(\left\{{\bf r}_j\right\}_{1,\ldots,N};R_C\right)=
\sum_{i>j=1}^N v(|{\bf r}_i-{\bf r}_j|)+\sum_{j=1}^N W(r_j;R_C),
\label{U_tot:eq}
\end{equation}
where $W(r;R_C)$ is given by Eqs.~(\ref{pot_liq:eq}) and (\ref{u_def:eq}).
Finding the ground-state configuration for the $N$-atom cluster
embedded in a liquid-like environment (at a given $T$ and $\rho$)
amounts to determining the global minimum of ${\cal U}$ 
with respect to variations of the positions of the $N$ atoms.
(Note that $T$ and $\rho$ only enter trough the external potential
$W(r;R_C)$, both explicitly (see Eq.~(\ref{pot_liq:eq}))
and through the state-dependence of the pair distribution function $g(r)$.)
It is worth stressing that, contrary to the case of an isolated cluster for 
which the radius of the cavity $R_C$ is merely fixed to avoid
evaporation of the atoms, $R_C$ becomes a relevant variable in a 
liquid-like environment: to preserve a realistic description of the liquid,
$R_C$ should adjust to global contractions or expansions of the $N$-atom
cluster, which should then be taken into account in the minimization
procedure. This point will be further discussed below.

To obtain the lowest energy minimum of ${\cal U}$, we have used
a slightly modified version of the basin-hopping algorithm introduced 
by Wales and co-workers~\cite{wales97,doye98}. 
The algorithm consists of a constant-temperature Monte-Carlo 
simulation performed with an acceptance criterion based not 
upon the energy of the proposed new configuration, but upon 
the energy of the closest minimum of the potential energy surface, 
obtained by a local minimization starting from that configuration. 
This algorithm turns out to be a very efficient method for exploring 
directly the minima of the potential energy surface, and it allows one 
to locate the ground-state with relatively little effort.     

Finally we have also considered other boundary conditions for the 
$N$-atom cluster:

{\em i) Free boundary conditions (isolated cluster).}-- 
This is the standard case studied in the literature.
It simply corresponds to the above situation in which the external 
potential is set to zero:
\begin{equation}
W(r;R_C)=0.
\label{pot_free:eq}
\end{equation}

{\em ii) Periodic boundary conditions (periodic replication
of the local cluster).}--
Each atom of the cluster now interacts also with the images
of the other atoms of the cluster
\begin{equation}
W({\bf r})=\frac{1}{2}\sum_{j \in {\cal C}} v(|{\bf r}-{\bf s}_j|),
\label{pot_pbc:eq}
\end{equation}
where ${\bf s}_j$ is the position of the image of atom $j$ selected 
through the minimum image criterion, i.e., among all possible images 
of atom $j$, only the closest is selected. We have used a cubic 
elementary cell.

{\em iii) Icosahedral-like boundary conditions 
(hypothetical non-frustrated system).}--
As mentioned in the Introduction, icosahedral order cannot be extended
to the entire space. One can however introduce icosahedral-like boundary
conditions by embedding the $13$-atom cluster in the center of a large
$147$-atom cluster with icosahedral $(I_h)$ symmetry in which the central
atom and its first layer have been removed.
(Recall that atomic clusters are characterized by ``magic numbers''
of atoms for which the global minimum has polytetrahedral 
symmetry~\cite{northby87,harris84}; in particular, icosahedral
symmetry is obtained for $13$, $55$, and $147$ atoms.)
The external potential is now written as 
\begin{equation}
W({\bf r})=\frac{1}{2}\sum_{k\in S_2^{ico}, S_3^{ico}}
v(|{\bf r}-{\bf r}_k|),
\label{pot_ico:eq}
\end{equation}
where $S_2^{ico}$ and $S_3^{ico}$ are the second and third shells of the 
$147$-atom icosahedron.
For completeness, we have also considered the $55$-atom
Mackay icosahedron, as well as the correction due to embedding the $55$-atom
or $147$-atom icosahedron in bulk liquid with the resulting
external potential treated at a mean-field level (see above).
\section{Global minimum of the Lennard-Jones clusters}
\label{results}
We specialize our investigation to the case of the Lennard-Jones pair potential
\begin{equation}
v_{LJ}(r)=4 \epsilon \left[
\left( \frac{\sigma}{r}\right)^{12}-
\left( \frac{\sigma}{r}\right)^{6} \right],
\label{LJ:eq}
\end{equation}
where $\epsilon$ and $2^{1/6} \sigma$ are the well-depth and the separation
at the minimum of the potential, respectively. In what follows we set 
$\sigma = \epsilon = 1$.

In order to evaluate the liquid-like external potential acting on the 
$N$-atom cluster, Eqs.~(\ref{pot_liq:eq}) and (\ref{u_def:eq}), 
and its derivatives, Eqs.~(\ref{der_RC:eq}) and (\ref{der_r:eq}),
one needs a model for the pair distribution function $g(r)$.
We use the $7$-parameter parametrization of 
Verlet's Molecular Dynamics simulation  data on the Lennard-Jones 
liquid~\cite{verlet68} proposed by Matteoli and 
Mansoori~\cite{matteoli95}:
\begin{equation}
g(y)=1+y^{-m}\left[g(d)-1-\lambda\right]+
\left[\frac{(y-1+\lambda)}{y}\right]
\exp{[-\alpha(y-1)]}\cos[\beta(y-1)],
\label{gdr_1:eq}
\end{equation}
for $m\ge 1, y\ge 1$, and
\begin{equation}
g(y)=g(d)\exp{[-\theta(y-1)^2]},
\label{gdr_2:eq}
\end{equation}
for $y<1$. Here $y=r/d$ is the dimensionless intermolecular distance
where $d=2^{1/6} \sigma$, and $h$, $m$, $\lambda$, $\alpha$, $\beta$, 
$\theta$, $g(d)$ are adjustable parameters. The terms $y^{-m}$ and 
$\exp{[-\theta(y-1)^2]}$ describe the decay of the first peak, while 
the term $\exp{[-\alpha(y-1)]}\cos[\beta(y-1)]$ provides the 
damped oscillations observed at larger distances~\cite{matteoli95}. 
We have taken the values $h=1.065$, $m=13.42$, $g(d)=2.830$, 
$\lambda=0.9310$, $\alpha=1.579$, $\beta=6.886$, 
$\theta=135.9$ that allow to reproduce the pair distribution function
for the liquid at $\rho=0.880$ and $T=1.095$ 
(in usual reduced Lennard-Jones units)~\cite{verlet68,matteoli95}.

\begin{figure}[p]
\centering
\includegraphics[width=0.80\textwidth]{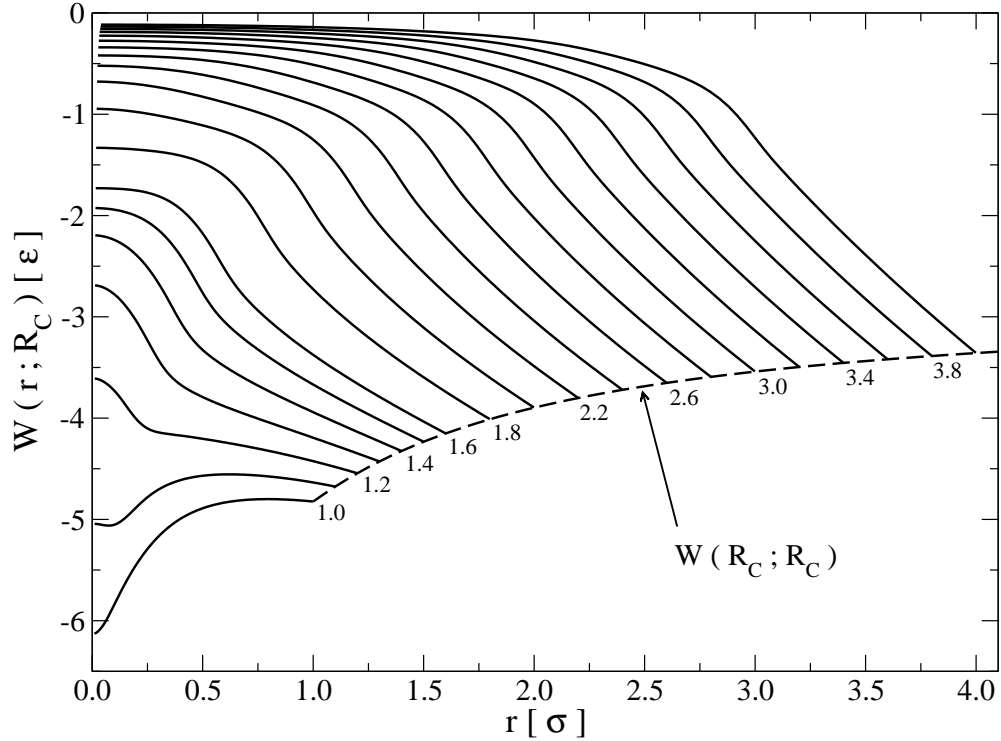}
\caption{Mean-field external potential as a function of 
the distance $r$ from the center of the cavity ; 
different curves are for different values of the radius 
of the cavity $R_C$. The dotted line is the value of the potential at the
surface of the cavity.}
\label{fig:meanfield}
\end{figure}

The resulting external potential $W(r;R_C)$ is shown in 
Fig.~\ref{fig:meanfield} for several values of the cavity radius $R_C$.
The shape of the $r$-dependence changes with $R_C$ so that no rescaling
of the curves is possible. For $R_C < 1.2\; \sigma$, the potential has a
minimum at $r=0$ because the central atom sits at the minimum -- or
very close to it -- of the pair interactions due to liquid atoms at the 
boundary of the cavity; this is no longer true for larger cavity radii,
and $W$ decreases monotonically with $r$, the most favorable position 
inside the cavity being at its edge where the attractive interaction 
due to the nearby liquid particles is the strongest. By construction, 
when $R_C\rightarrow 0$, $W(r=0,R_C)$ becomes equal to the total 
potential energy of the Lennard-Jones liquid at the considered
state point, $E\simeq -5.7\; \epsilon$, whereas when $R_C\rightarrow \infty$,
$W(r=R_C,R_C)$ is equal to half this energy.

As we discussed it above, the radius of the cavity $R_C$ must be adjusted 
to global contractions or expansions of the cluster.
A reasonable way to implement this is to take at each minimization step,
i. e., for each configuration of the $N$ atoms,
\begin{equation}
R_C=r_{max}+\mu \;\sigma,
\label{RC_def:eq}
\end{equation}
where $r_{max}$ is the distance of the outermost atom from the
center of the cavity, and $\mu$ is a constant chosen to account for the
fact that repulsive interactions between atoms make very unlikely
the presence of ``bulk liquid'' atoms when their centers are too close to 
those of the cavity atoms; we have taken $\mu=0.5$, but we have checked 
that the results are independent of the actual value, in runs with 
different values of $\mu$ between $0.1$ and $1.0$.

\begin{figure}[p]
\centering
\includegraphics[width=0.50\textwidth]{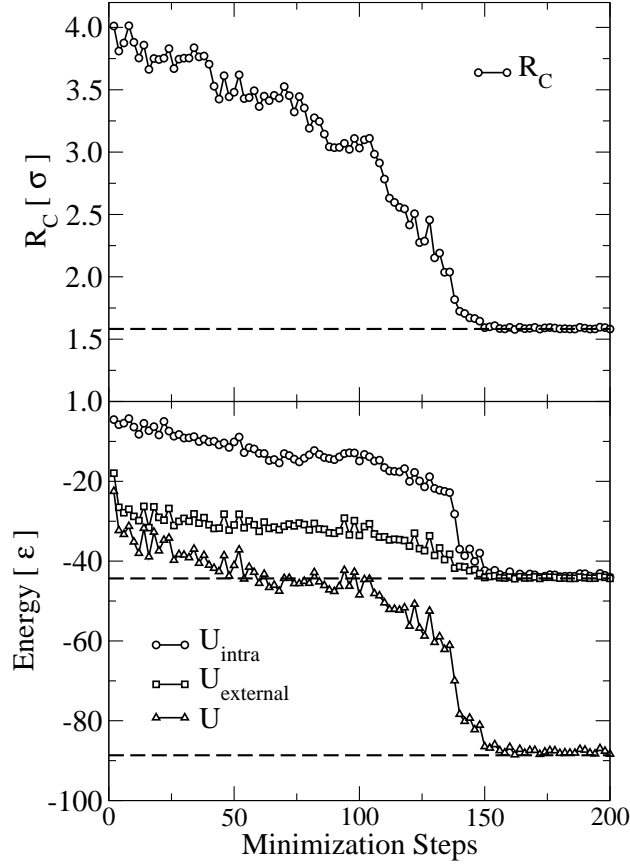}
\caption{Evolution of the cavity radius (Top) and of the cluster energy,
together with its intra-cluster and external-field contributions (Bottom), 
during a minimization run starting from a random $13$-atom Lennard-Jones
cluster with mean-field liquid-like environment.}
\label{fig:minimization}
\end{figure}

A typical minimization run for a $13$-atom cluster in the presence of
a mean-field liquid-like environment is shown in 
Fig.~\ref{fig:minimization}, where we have plotted the evolution of the 
energy and its intra-cluster and external-field contributions (bottom),
together with the evolution of the cavity radius $R_C$ (top).
During the optimization run that starts from a random configuration
of $13$ atoms, the structure of the cluster becomes more and more
compact, and its energy decreases. The final optimized configuration
is found to have icosahedral symmetry. Both the final cluster
radius, $r_{max} \simeq 1.08\; \sigma$, and the final intra-cluster
energy, $U_{intra} \simeq -44.327 \;\epsilon$ are identical to those 
found for the ground-state of the isolated $13$-atom cluster~\cite{northby87}. 
The total energy of the icosahedral cluster in the presence of a liquid-like 
environment is however much lower (by almost a factor two) because of the 
external field that compensates for the deficiency of nearest neighbors 
of the $12$ surface atoms.
We have repeated the procedure for different starting random configurations
and we have always obtained the icosahedron with $r_{max} \simeq 1.08\; \sigma$
as the global minimum. The same is true for a whole range of liquid density
$\rho$ and temperature $T$ ($0.65\le \rho \le0.88, 0.6\le T \le 3.6$).

\begin{figure}[p]
\centering
\includegraphics[width=0.80\textwidth]{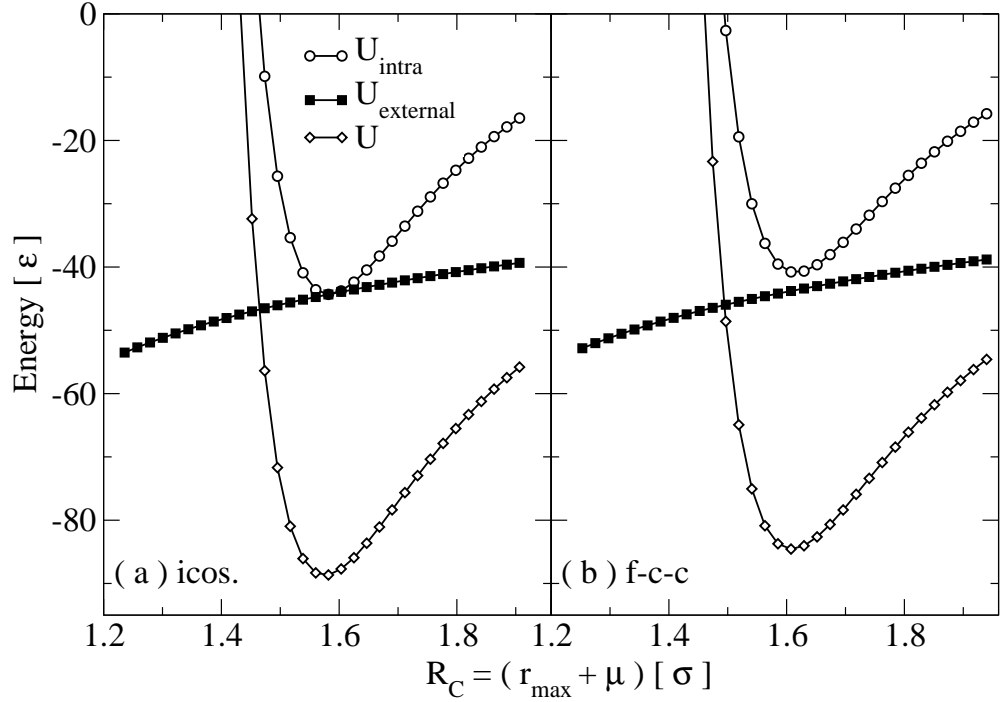}
\caption{Energy of a $13$-atom cluster with mean-field liquid-like 
environment as a function of $R_C=r_{max}+\mu\; \sigma$ (diamonds): 
{\em (a)} icosahedral $(I_h)$ cluster, {\em (b)} f-c-c $(O_h)$ cluster. 
We show separately the contributions of the intra-cluster interaction 
energy (circles) and that of the external field (squares).}
\label{fig:rmax}
\end{figure}

In addition to carrying out a global optimization procedure,
one may also make a calculation in the spirit of Frank's pioneering 
work~\cite{frank52}. We have considered two potential candidates
for the ground-state configuration of the $13$-atom cluster,
namely the icosahedron and the cuboctahedral cluster with $O_h$ symmetry
that is associated with the face-centered-cubic close-packed
lattice and we have compared their energies in a liquid-like
environment. The energies of the two, $I_h$-symmetric and $O_h$-symmetric,
clusters are shown in Fig.~\ref{fig:rmax} as a function of the cavity
radius $R_C$ (or, equivalently, as a function of $r_{max}$, a unique 
distance being enough to fully determine the whole cluster once the symmetry,
$I_h$ or $O_h$, is chosen.). The intra-cluster contribution to the energy
has a minimum for $r_{max} \simeq 1.08\; \sigma$ for the icosahedron,
and $r_{max} \simeq 1.10\; \sigma$ for the $O_h$ cluster, whereas the 
external-field contribution monotonously increases with $R_C$ in both cases.
One then finds that the icosahedral cluster has a minimum total energy
for $r_{max} \simeq 1.08\; \sigma$ and that this energy is $4.8 \%$ lower than
the lowest energy found for the $O_h$ cluster 
when $r_{max} \simeq 1.10\; \sigma$.
When compared to Frank's result $(8.4\%)$, the relative energy difference
between the two types of clusters is thus not drastically modified: 
a mean-field liquid-like environment only slightly reduces the relative
stability of the icosahedral order. 
Finally, as a mere check of our global optimization procedure, we have 
verified that the icosahedral ground-state found here is identical to 
that discussed above.

The influence of the boundary conditions on the ground-state of a $13$-atom
cluster can be investigated by using again the optimization algorithm
(always starting with a random initial configuration) with the appropriate
conditions, free, periodic, and icosahedral-like, described in the previous
Section. As already well-known, the ground-state of the isolated cluster
is an icosahedron and, as anticipated, that of the cluster in the presence
of icosahedral-like boundary conditions is also an icosahedron.
The ground-state energies are shown in Table~\ref{table:energies},
and can be compared to that of the icosahedral cluster in a mean-field
liquid-like environment. Not surprisingly, this latter is much lower
than that of the isolated cluster (see above), but it is higher
than that of icosahedra embedded in larger icosahedral structures.

We note on passing that, in the case of an icosahedral-like environment,
the change in the structure of the $13$-atom cluster from random to 
icosahedral during the optimization run starts from the outside and propagates 
inward. This is reminiscent of what has been observed in the simulation
of gold nanoclusters~\cite{nam02}. There, it has been found that, just after
freezing, ordered nanosurfaces with five-fold rotational symmetry
are formed, while interior atoms remain in a disordered state.
On lowering the temperature, the crystallization of the interior
atoms proceeds from the surface toward the core region, eventually
producing an icosahedral structure~\cite{nam02}.
This is at variance with the classical picture of homogeneous
nucleation and rather represents a surface-induced (heterogeneous)
crystallization.

\begin{figure}
\centering
\includegraphics[width=0.80\textwidth]{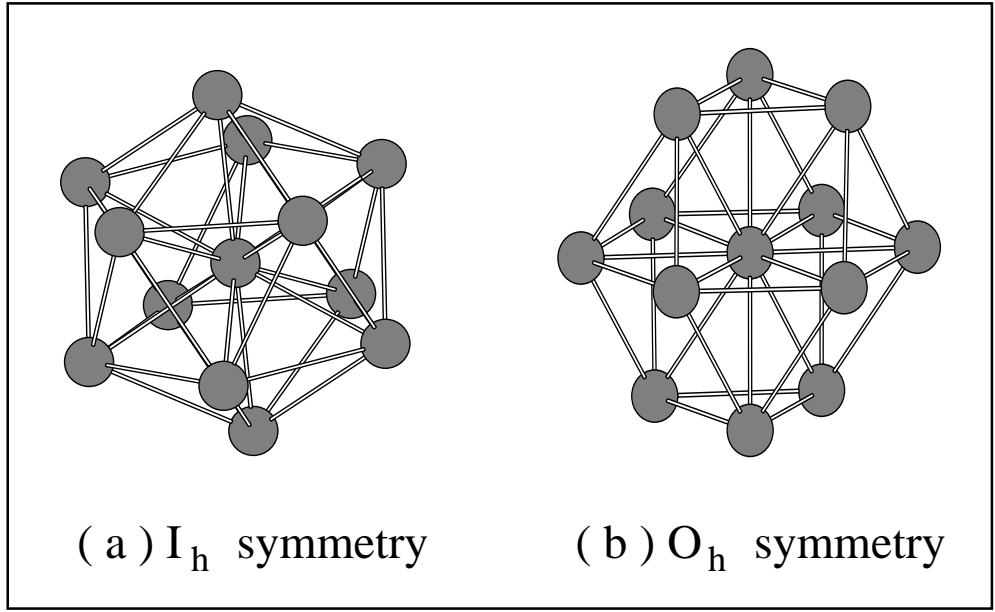}
\caption{Structure of the icosahedral {\em (a)} 
and cuboctahedral {\em (b)} $13$-atom clusters.}
\label{fig:groundstates}
\end{figure}

Periodic boundary conditions lead to a quite different picture.
Since icosahedra cannot tile space by periodic replication,
such conditions should favor the symmetries that allow a complete filling
of space with true long-range order.
It is indeed what we have found: the global minimum is then a cuboctahedral
cluster with $O_h$ symmetry, that leads to a face-centered-cubic 
close-packed lattice when 
periodically replicated (see Fig.~\ref{fig:groundstates}); the corresponding
ground-state energy is given in Table~\ref{table:energies}, and it is found
lower than that of an isolated icosahedron --- because of the lack of
neighbors already mentioned for this latter case --- but higher than icosahedra
in either liquid-like or icosahedral-like environment.

\begin{figure}[p]
\centering
\includegraphics[width=0.70\textwidth]{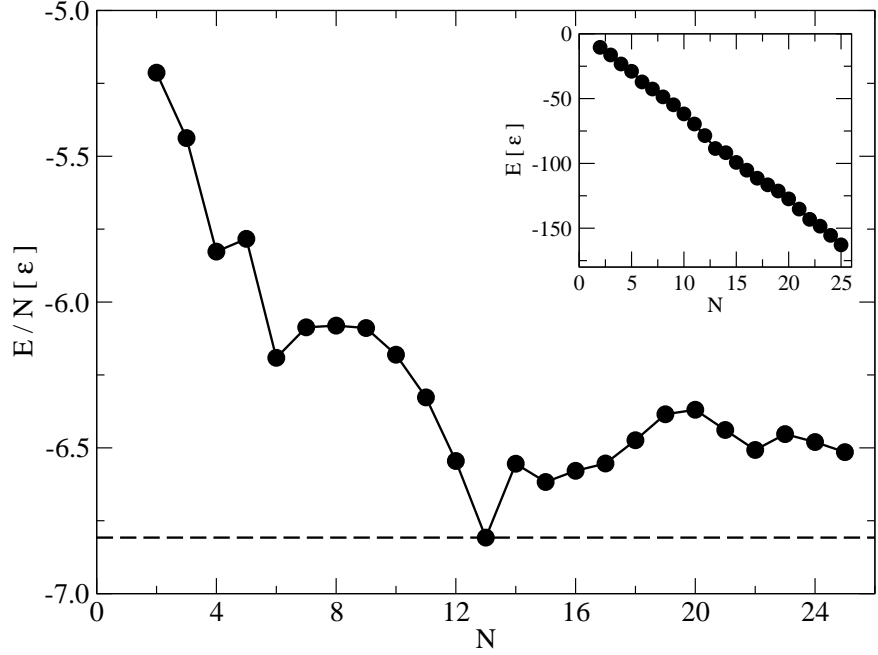}
\caption{Energy per atom of the ground-state of $N$-atom 
clusters with mean-field liquid-like environment. {\em Inset:} total
energy as a function of $N$.}
\label{fig:energies}
\end{figure}

Finally, we have considered the effect of varying the number of atoms 
present in the cluster. The main motivation for this study is to check
that without a-priori knowledge of the preferred local structure in the presence
of a mean-field liquid-like environment, hence of the number of atoms
involved in this structure, the global minimization method will help
select the proper preferred configuration. This will be important when
considering molecular liquids.
We have thus studied the global minima of $N$-atom clusters with $N$
ranging from $2$ to $23$. As shown in Fig.~\ref{fig:energies}, the energy
per-atom, the only relevant quantity for comparing local structures in a 
liquid-like environment, is lowest for the $N=13$ icosahedral ground-state.
For small $N$, the mean-field description of the liquid environment is
probably too crude to give sensible results, but it is nonetheless significant 
that for a large range of $N$, the icosahedral cluster is properly selected 
as the locally preferred structure of the Lennard-Jones liquid. 
\section{Conclusion}
\label{conclusion}
Local icosahedral order has been found both in bulk condensed phases
and in clusters formed by spherical particles.
The connection between bulk and cluster studies is however obscured
by two facts: first, geometrical frustration strongly hinders
the spatial extension of local icosahedral order and distorts the
local icosahedra in bulk conditions; second, the energetics of isolated
clusters is partly determined by surface effects that are of course absent
in the bulk. In this work we have tried to bypass those problems. 
We have proposed to determine the locally preferred structure
of a liquid by finding the ground-state configuration of $N$-particle 
clusters embedded in a liquid-like continuum, characterized by the proper 
density and pair distribution function of the bulk liquid at the chosen 
thermodynamic state point.
This mean-field-like procedure minimizes the surface effects without 
introducing full blown geometrical frustration.

In terms of potential energy surface, we have therefore introduced
an effective energy surface that contains the usual intra-cluster
potential energy contribution plus an external field that accounts 
for the interaction with the outside liquid at a mean-field level. 
By a global optimization algorithm we have then located the 
lowest-energy minimum of the effective energy surface. For Lennard-Jones 
pair interactions, we have found that the locally preferred structure
is indeed an icosahedron, the effect of the liquid-like environment
being to only slightly reduce the relative stability of
the icosahedral structure when compared
to Frank's calculation for an isolated cluster~\cite{frank52}.
We have also shown the importance of the boundary conditions used
for the cluster: whereas icosahedral-like boundary conditions stabilize
even more the local icosahedral cluster, periodic boundary conditions
make the cuboctahedral cluster more stable, a consequence of geometrical frustration
that prevents tiling of space by icosahedra, and therefore favors 
long-range order associated with face-centered cubic or hexagonal
close-packed lattices.

The present findings for the Lennard-Jones liquid suggest that
the proposed method could be efficient as well for determining the 
locally preferred structure of molecular liquids, in cases where
both translational and rotational degrees of freedom are involved,
and a-priori knowledge about the putative local order is scarce. 
\begin{acknowledgments}
We thank D.~J.~Wales and F.~Sciortino for fruitful comments.
\end{acknowledgments}
%
%
\appendix*
\section{Mean-field liquid external potential}
\label{calcul:app}
Here we give some details about the calculation of the mean-field
liquid external potential $W(r;R_C)$ discussed in Sect.~\ref{method}.

\begin{figure}[p]
\centering
\includegraphics[width=0.60\textwidth]{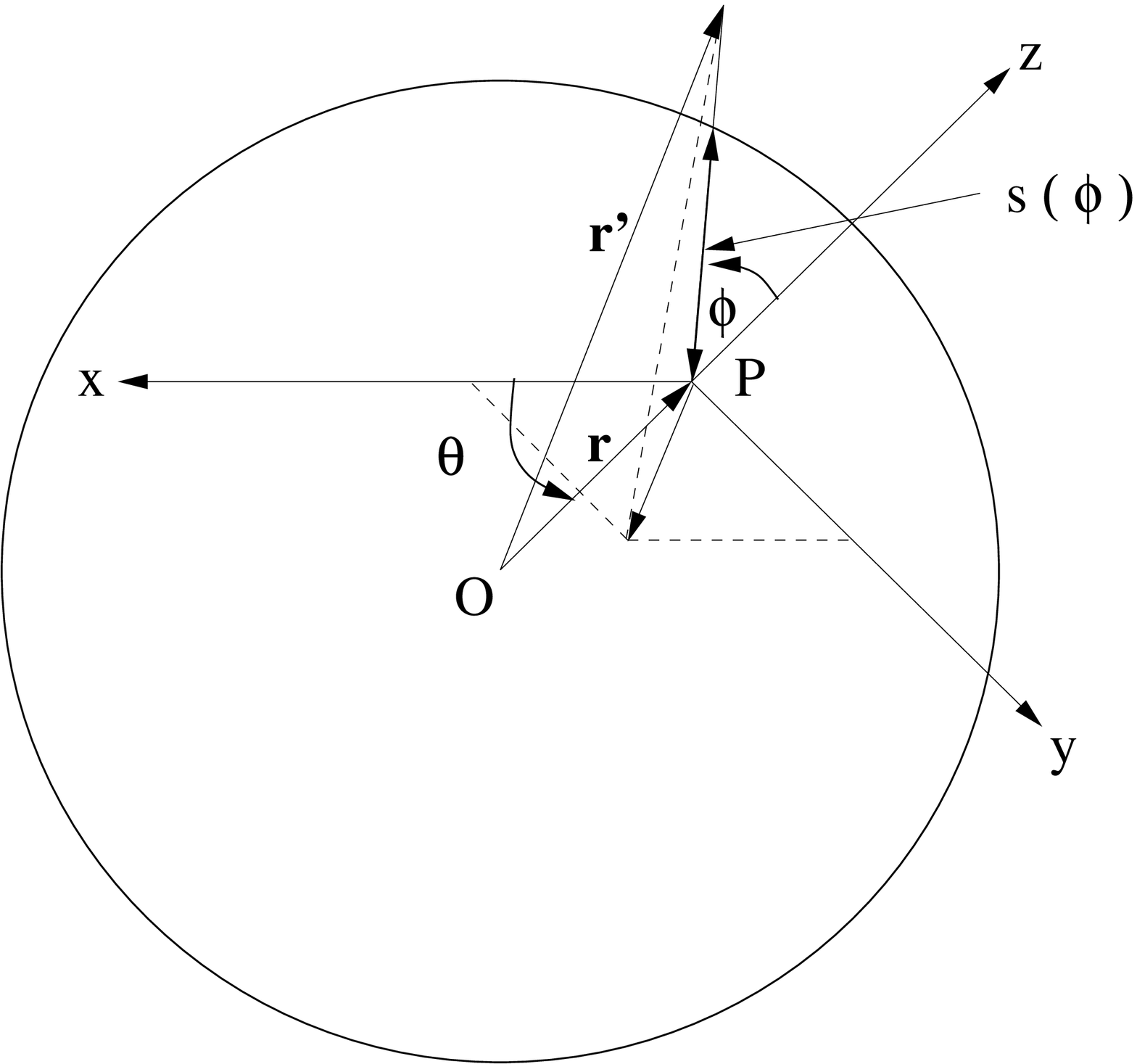}
\caption{Geometry used for the calculation of the external potential 
acting on an atom located at point $P$ in a cavity of radius $R_C$.}
\label{fig:geometry}
\end{figure}

From Eq.~(\ref{meanfield:eq}) the external potential felt by an atom
at ${\bf r}$, where the origin of the coordinates is chosen as the 
center of the cavity, can be written as
\begin{equation}
W(r,R_C)=\frac{\rho}{2}\int_{|{\bf x}+{\bf r}|>R_C} d^3 x\; g(x) v(x),
\label{pot_liq0:eqapp}
\end{equation}
where we have changed the integration variable from ${\bf r'}$ to 
${\bf x}={\bf r}-{\bf r'}$.
We now rotate the reference system such that  ${\bf r}/r={\bf \hat z}$ 
and we translate it so that the new origin is at $P$, the position of the
center of the atom under consideration.
The new geometry of the problem is shown in Fig.~\ref{fig:geometry}.
One can easily convince oneself that
\begin{equation}
W(r;R_C)=\frac{\rho}{2}\int_0^\pi d\phi\; \sin \phi 
\int_0^{2 \pi} d\theta
\int_{s(\phi)}^{\infty} dx\; x^2 g(x) v(x).
\label{pot_liq1:eqapp}
\end{equation}
Given a point inside the cavity, $s$ is its distance 
from the surface in the direction ${\bf x}$.
It is simple to check that $s$ is a function of $\phi$ only and is solution 
of the quadratic equation
\begin{equation}
s^2+(2 r \cos \phi) s+(r^2-R_C)^2=0;
\label{rho_phi:eqapp}
\end{equation}
the correct solution is
\begin{equation}
s(\phi)=-r \cos\phi+\sqrt{r^2\cos^2\phi+(R_C^2-r^2)}.
\label{rho:eqapp}
\end{equation}
Introducing the variable $u=\cos \phi$ we obtain
\begin{equation}
W(r;R_C)=\pi \rho\int_{-1}^{1} du \int_{s(u)}^{\infty} 
dx\; x^2 g(x) v(x),
\label{pot_liq2:eqapp}
\end{equation}
with
\begin{equation}
s(u)=-r u +\sqrt{r^2(u^2-1)+R_C^2}.
\label{rhou:eqapp}
\end{equation}
$s(u)$ is such that $s (-1) = R_C+r$, and $s (1) = R_C-r$.
We now change variable from $u$ to $s$, and integrate by parts so that
\begin{eqnarray}
W(r;R_C)&=&\pi \rho\int_{R_C+r}^{R_C-r} ds \;\frac{du}{ds}
\int_\rho^\infty dx\; x^2 g(x) v(x)\nonumber\\
&=&\pi \rho \left\{\left[u(s)\int_\rho^\infty dx\; x^2 g(x) v(x)\right]
^{R_C-r}_{R_C+r}
-\int_{R_C-r}^{R_C+r} ds\; s^2 u(s) g(s) v(s)
\right\},
\label{pot_liq3:eqapp}
\end{eqnarray}
where
\begin{equation}
u(s)=\frac{R_C^2-r^2-s^2}{2 r s},
\label{urho:eqapp}
\end{equation}
and $u(s=R_C-r)=-u(s=R_C+r)=1$.
Substituting Eq.~(\ref{urho:eqapp}) in Eq.~(\ref{pot_liq3:eqapp}) and 
rearranging we obtain Eqs.~(\ref{pot_liq:eq}) and~(\ref{u_def:eq}). 
\newpage

\begin{table}[p]
\centering
\begin{tabular}{|l||l|c|}
\hline
Boundary Conditions & Energy $[ \epsilon ]$ & Point Group\\\hline\hline
Free                      & $-44.327$  & $I_h$\\
Periodic                  & $-52.745$  & $O_h$\\
Liquid-like               & $-88.490$  & $I_h$\\
Icos. (1 lay.)            & $-92.980$  & $I_h$\\
Icos. (2 lay.s)           & $-100.142$ & $I_h$\\
Icos. (1 lay.) \& Liquid  & $-101.395$ & $I_h$\\
Icos. (2 lay.s) \& Liquid & $-102.208$ & $I_h$\\
\hline
\end{tabular}
\vspace{0.5cm}
\caption{Ground-state energy and symmetry of the $13$-atom cluster
with various boundary conditions. The icosahedral-like
conditions correspond to the $55$-atom (1 layer)
and $147$-atom (2 layers) icosahedral clusters, and the same 
structures embedded in a mean-field liquid-like environment.}
\label{table:energies}
\end{table}
\vspace{5.0cm}

\begin{thebibliography}{99}
%
\bibitem{sadoc99}
J.~F.~Sadoc and R.~Mosseri, {\em ``Geometric Frustration''},
(Cambridge University Press, Cambridge, 1999).
%
\bibitem{nelson89}
D.~R.~Nelson, Phys. Rev. B {\bf 28}, 5515 (1983);
D.~R.~Nelson and F.~Spaepen, Solid State Physics {\bf 42}, 1 (1989);
D.~R.~Nelson, {\em ``Defects and Geometry in Condensed Matter Physics''},
(Cambridge University Press, Cambridge, 2001).
%
\bibitem{kivelson95}
D.~Kivelson, S.~A.~Kivelson, X.~Zhao, Z.~Nussinov, and G.~Tarjus,
Physica A {\bf 219}, 27 (1995).
%
\bibitem{frank52}
F.~C.~Frank, Proc. Roy. Soc. A {\bf 215}, 43 (1952).
%
\bibitem{filipponi99}
A.~Filipponi, A.~ Di Cicco, and S.~De Panfilis, 
Phys. Rev. Lett. {\bf 83}, 560 (1999).
%
\bibitem{reichert00}
H.~Reichert, O.~Klein, H.~Dosch, M.~Denk, V.~Honkim\"aki, T.~Lippmann, and 
G.~Reiter, Nature (London) {\bf 408}, 839 (2000).
%
\bibitem{shenk02}
T.~Shenk, D.~Holland-Moritz, V.~Simonet, R.~Bellissent, and D.~M.~Herlach,
Phys. Rev. Lett. {\bf 89}, 075507 (2002).
%
\bibitem{steinhardt83}
P.~J.~Steinhardt, D.~R.~Nelson, and M.~Ronchetti,
Phys. Rev. B {\bf 28}, 784 (1983).
%
\bibitem{jonsson88}
H.~J\'onsson and H.~C.~Andersen,
Phys. Rev. Lett. {\bf 60}, 2295 (1988).
%
\bibitem{kondo90}
T.~Kondo, K.~Tsumuraya, and M.~S.~Watanabe, 
J. Chem. Phys. {\bf 93}, 5182 (1990); 
T.~Kondo and K.~Tsumuraya, ibid. 
{\bf 94}, 8220 (1991). 
%
\bibitem{yonezawa91}
F.~Yonezawa, Solid State Physics {\bf 45}, 179 (1991).
%
\bibitem{shumway95}
S.~L.~Shumway, A.~S.~Clarke, and H.~J\'onsson, 
J. Chem. Phys. {\bf 102}, 1796 (1995).
%
\bibitem{jund97}
P.~Jund, D.~Caprion, and R.~Jullien, 
Europhys. Lett. {\bf 37}, 547 (1997).
%
\bibitem{dzugutov02}
M.~Dzugutov, S.~I.~Simdyankin, and F.~H.~M.~Zetterling,
Phys. Rev. Lett. {\bf 89}, 195701 (2002).
%
\bibitem{doye03}
J.~P.~K. Doye, D.~J.~Wales, F.~H.~M.~Zetterling, and M.~Dzugutov,
J. Chem. Phys. {\bf 118}, 2792 (2003).
%
\bibitem{doye01}
J.~P.~K. Doye, D.~J.~Wales, and S.~I.~Simdyankin,
J. Chem. Soc. Faraday Discuss. {\bf 118}, 159 (2001).
%
\bibitem{sachdev92}
S.~Sachdev, in {\em ``Bond-Orientational Order in Condensed Matter Systems''},
K.~J.~Strandburg Ed. (Springer-Verlag, New York, 1992), p. 255.
%
\bibitem{wales97}
D.~J.~Wales and J.~P.~K.~Doye, J. Phys. Chem. A {\bf 101}, 5111 (1997).
%
\bibitem{doye98}
J.~P.~K.~Doye and D.~J.~Wales, 
Phys. Rev. Lett. {\bf 80}, 1357 (1998).
%
\bibitem{northby87}
J.~A.~Northby, J. Chem. Phys. {\bf 87}, 6166 (1987).
%
\bibitem{harris84}
I.~A.~Harris, R.~S.~Kidwell, and J.~A.~Northby,
Phys. Rev. Lett. {\bf 53}, 2390 (1984).
%
\bibitem{verlet68}
L.~Verlet, Phys. Rev. {\bf 165}, 201 (1968).
%
\bibitem{matteoli95}
E.~Matteoli and G.~A.~Mansoori,
J. Chem. Phys. {\bf 103}, 4672 (1995).
%
\bibitem{nam02}
H.~-S.~Nam, N.~M.~Hwang, B.~D.~Yu, and J.~-K.~Yoon,
Phys. Rev. Lett. {\bf 89}, 275502 (2002).
%
\end{thebibliography}
\end{document}